\documentstyle[12pt,psfig]{article}
\def\ra{\rightarrow}

\def\be{\begin{equation}}
\def\ee{\end{equation}}
\def\bea{\begin{eqnarray}}
\def\eea{\end{eqnarray}}
\def\la{\langle}
\def\ra{\rangle}
\def\dis{\displaystyle}

\title{Higher Order Hybrid Monte Carlo at Finite Temperature}

\author{Tetsuya Takaishi
\\
{\small\sl Hiroshima University of Economics, Hiroshima 731-0192, Japan}\\
}

\date{}

\begin{document}
\maketitle
\abstract{
The standard hybrid Monte Carlo algorithm uses 
the second order integrator at the molecular dynamics step.
This choice of the integrator is not always the best.
Using the Wilson fermion action,
we study the performance of the hybrid Monte Carlo algorithm for lattice QCD 
with higher order integrators in both zero and finite temperature phases and
find that in the finite temperature phase the performance 
of the algorithm can be raised by use of the 4th order integrator. 
\\
\\
\\

}
\newpage
\section{Introduction}
In lattice QCD the hybrid Monte Carlo (HMC) algorithm \cite{HMCA} is widely used 
for simulations of even-flavor of quarks\footnote{Odd-flavor simulations of 
the HMC algorithm are also possible by modifying the Hamiltonian \cite{ODDHMCA}.}.
These simulations are usually difficult tasks, especially at small quark masses
where the computational cost of the matrix solver 
which is the most time consuming part of the HMC algorithm grows.
In order to obtain reliable results within limited computer resources  
it is important to find an efficient way to implement 
the HMC algorithm so that 
the total computational cost of the algorithm is minimized \cite{Recent}.

The basic idea of the HMC is a combination of 
1) molecular dynamics and 2) Metropolis test.
Usually the second order leapfrog integrator 
is used at the molecular dynamics (MD) step. 
The integrator causes ${\cal O}(\Delta t^3)$ integration errors,
where $\Delta t$ denotes the step-size.
Due to the integration errors the Hamiltonian is not conserved. 
The errors introduced by this integrator have to be removed by 
the Metropolis test, i.e. accept the new configuration with a probability  
$\sim \min(1,\exp(-\Delta H))$ where $\Delta H=H_{new}-H_{old}$ is an energy difference  
between the starting Hamiltonian $H_{old}$ and the new Hamiltonian $H_{new}$ at the end of the trajectory. 

The acceptance at the Metropolis step depends on the magnitude of the energy difference 
induced by the integration errors. 
If a higher order integrator is used 
at the MD step, the integration errors can be reduced.
Therefore one may easily imagine that the performance of 
the HMC increases with the higher order integrator.  
However this is not always true since the higher order integrator has more arithmetic operations than
the lower one and this might decrease the performance. 
The total performance should be measured 
by taking account of two effects: acceptance and the number of arithmetic operations.
In Ref.\cite{Takaishi} the performance of the higher order integrator at zero temperature ($\beta=0$)
was studied systematically and it turned out that for the simulation parameters 
used for the current large-scale simulations with the Wilson fermion action 
the 2nd order integrator is the best one.
The main reason why the higher order integrators are not so effective
is that the energy difference caused by the errors of the higher order integrator increases 
more rapidly than that of the lower one as the quark mass decreases. 
In Ref.\cite{Accept} it is shown that at finite temperature 
the quark mass dependence of the energy difference is small. 
If so, the conclusion of Ref.\cite{Takaishi} might change at finite temperature.
In this letter we test higher order integrators at  finite temperature and
demonstrate that they can actually perform  better than  the lower order.

\section{Higher order integrator}
In this section we define higher order integrators studied here.
Our definition is parallel to that of Ref.\cite{Takaishi}. 
Let $H$ be a Hamiltonian given by
\be
H=\frac12 p^2 +S(q)
\ee
where $q=(q_1,q_2,...)$ and $p=(p_1,p_2,...)$
are coordinate variables and conjugate momenta respectively.
$S(q)$ is a potential term of the system considered.
For the lattice QCD, $q$ correspond to link variables and
$S(q)$ consists of gauge and fermion actions.

In the MD step we solve Hamilton's equations of motion,
\begin{equation}
\left\{
\begin{array}{ll}
\dis\frac{dq_i}{dt} & = \dis\frac{\partial H}{\partial p_i}\\
\dis\frac{dp_i}{dt} &= \dis-\frac{\partial H}{\partial q_i}.
\end{array}
\right.
\end{equation}
In general these equations are not solvable analytically.
Introducing a step-size $\Delta t$, the discretized version of 
the equations are solved.
In the conventional HMC simulations the 2nd order leapfrog 
scheme, which causes ${\cal O}(\Delta t^3)$ step-size error, is used
to solve the equations. 
This scheme is written as
\begin{equation}
\left\{
\begin{array}{ll}
q(t+\frac{\Delta t}{2}) & =  q(t)+\frac{\Delta t}{2} p(t) \\
p(t+\Delta t) & =  p(t) -\Delta t \dis\frac{\partial S(q(t+\frac{\Delta
t}{2}))}{\partial q} \\
q(t+\Delta t)& = q(t+\frac{\Delta t}{2}) +\frac{\Delta t}{2} p(t+\Delta t).
\end{array}
\right.
\label{2ND}
\end{equation}
Eq.(\ref{2ND}) forms an elementary MD step. 
This elementary MD step is performed repeatedly 
$N$ times. The trajectory length $\tau$ is given by $\tau=N\times \Delta t$.

Any integrator which satisfies two conditions: 1)time reversible and 2)area preserving 
can be used for the MD step of the HMC. 
These conditions are needed to satisfy the detailed balance.
When we use the Lie algebraic formalism \cite{SW,HOHMC,Yoshida,Suzuki} we can easily 
construct higher order integrators which satisfy the above conditions. 
From the Lie algebraic formalism 
we find that higher order integrators can be constructed 
by combining lower order integrators.
Let $G_{2nd}(\Delta t)$ be an elementary MD step of the 2nd order integrator 
with a step-size $\Delta t$.
The 4th order integrator $G_{4th}(\Delta t)$ is constructed 
from a product of $three$ 2nd order integrators as \cite{SW,HOHMC,Yoshida,Suzuki,4th} 
\begin{equation}
G_{4th}(\Delta t) = G_{2nd}(a_1 \Delta t) G_{2nd}(a_2 \Delta t) G_{2nd}(a_1
\Delta t),
\label{4TH}
\end{equation}
where the coefficients $a_i$ are given by
\begin{equation}
a_1 = \frac{1}{2-2^{1/3}} ,
\end{equation}
\begin{equation}
a_2 = - \frac{2^{1/3}}{2-2^{1/3}}.
\end{equation}
Eq.(\ref{4TH}) means that there are three elementary MD steps:
(i) first we integrate the equations by eq.(\ref{2ND}) with  a step-size of $a_1 \Delta t$,
(ii) then proceed the 2nd order integration with a step-size of $a_2 \Delta t$,
(iii) finally integrate the equations with a step-size of $a_1 \Delta t$. 
After performing these three elementary MD steps sequentially we obtain the 4th order integrator 
with the step-size $\Delta t$. 
This construction scheme can be generalized to the  higher $even$-order  integrators.
The $(2k+2)$-th order integrator is given recursively by
\begin{equation}
G_{2k+2}(\Delta t) = G_{2k}(b_1 \Delta t) G_{2k}(b_2 \Delta t) G_{2k}(b_1
\Delta t),
\label{rec}
\end{equation}
where the coefficient $b_i$ are
\begin{equation}
b_1 = \frac{1}{2-2^{1/(2k+1)}}
\end{equation}
\begin{equation}
b_2 = - \frac{2^{1/(2k+1)}}{2-2^{1/(2k+1)}}.
\end{equation}
Compared with the 2nd order integrator, the computational cost of 
the $n$-th order one constructed from eq.(\ref{rec}) grows as  $3^{n/2-1}$. 
For instance the 6th order integrator has 9 times more arithmetic operations than
those of the 2nd order one. 
Yoshida \cite{Yoshida} found $three$ parameter sets of 6th order integrators 
with less arithmetic operations (Table.\ref{Table1}).
Yoshida's 6th order integrators consist of $seven$ $G_{2nd}$'s as
\begin{eqnarray}
G_{6th}(\Delta t)&=&G_{2nd}(w_3 \Delta t) G_{2nd}(w_2 \Delta t) G_{2nd}(w_1
\Delta t)G_{2nd}(w_0 \Delta t) \nonumber \\
& & \times G_{2nd}(w_1 \Delta t) G_{2nd}(w_2 \Delta t) G_{2nd}(w_3 \Delta
t).
\label{6th}
\end{eqnarray}
In Ref.\cite{Takaishi} these 6th order integrators were examined and one of three parameter sets, 
denoted by $Y1$, was 
found to give smaller integration errors than those of others.
In this study the parameter set $Y1$ ( see Table.\ref{Table1} ) is used for the 6th order integrator. 

\begin{table}[h]
\begin{center}
\begin{tabular}{c|ccc}  \hline
    & Y1 & Y2& Y3 \\ \hline
$w_1$ & -0.117767998417887e-1 & -0.213228522200144e+1 &
0.152886228424922e-2 \\
$w_2$ & 0.235573213359357e+0  & 0.426068187079180e-2
  & -0.214403531630539e+1 \\
$w_3$ & 0.784513610477560e+0  & 0.143984816797678e+1   &
0.144778256239930e+1  \\ \hline

\end{tabular}
\caption{\small Parameter sets (Y1-Y3) of the 6th order integrators
by Yoshida {\cite{Yoshida}}.
$w_0$ is given by $w_0=1-w_1-w_2-w_3$.
}
\label{Table1}
\end{center}
\end{table}

\section{Efficiency of the HMC algorithm}
In order to compare among various integrators 
one needs a criterion which ranks integrators. 
Following Ref.\cite{Takaishi} we utilize the efficiency function $E_{ff}$
constructed from a product of acceptance $P_{acc}$ and step-size $\Delta t$:
\be
E_{ff}=P_{acc}\Delta t.
\label{efficiency}
\ee
This function has one maximum at a certain step-size
which we denote $\Delta t_{opt}$.
Using the energy difference $\Delta H$ 
the acceptance is expressed by \cite{Accept} 
\begin{eqnarray}
\la P_{acc}\ra & = & {\rm erfc}(\frac{1}{2}\la \Delta
H\ra^{1/2}),
\label{acc}
\end{eqnarray}
where erfc is the complementary error function.
In stead of using eq.(\ref{acc}),
when $\la\Delta H\ra$ is small, we may  use 
\begin{equation}
\la P_{acc}\ra=\exp(-\frac{2}{\sqrt{\pi}}\la \frac{1}{8}\Delta
H^2\ra^{1/2}).
\label{Papprox}
\end{equation}
Although mathematically speaking eq.(\ref{Papprox}) is valid only for $\langle \Delta H^2\rangle^{1/2} \ll 1$, 
the numerical study \cite{Takaishi} shows that  eq.(\ref{Papprox}) 
approximates the acceptance quit well for  
$\langle \Delta H^2\rangle^{1/2} \leq 3$ which corresponds to $\la P_{acc}\ra\geq 20\%$. 
This is enough for our purpose since typically the acceptance of the HMC 
is taken to be $\la P_{acc}\ra\geq 50\%$.

In the lowest order of $\Delta t$, $\langle \Delta H^2\rangle^{1/2}$ of the $n$-th order integrator 
is expressed as \cite{Takaishi,Accept}
\begin{equation}
\langle \Delta H^2\rangle^{1/2} = C_n V^{1/2} \Delta t^n +
{\cal O} (\Delta t^{n+1}),
\label{dH2}
\end{equation}
where $V$ is volume of the system and $C_n$ is a Hamiltonian dependent coefficient. 

Substituting eq.(\ref{dH2}) into eq.(\ref{Papprox}) one obtains
\begin{equation}
\langle P_{acc}\rangle = \exp(-\widetilde{C}_n V^{\frac{1}{2}} \Delta t^n),
\label{Accaprox}
\end{equation}
where $\widetilde{C}_n \equiv C_n/\sqrt{2\pi}$.
If one uses eq.(\ref{Accaprox}) for eq.(\ref{efficiency}) 
one can easily obtain the optimal step-size which gives the maximum efficiency:
\begin{equation}
\Delta t_{opt} = \sqrt[n]{\frac{1}{n \widetilde{C}_n V^{\frac{1}{2}}}  }.
\label{DTopt}
\end{equation}
Furthermore substituting $\Delta t_{opt}$ to eq.(\ref{Accaprox}) one obtains 
the optimal acceptance\footnote{A recent study \cite{CPC} shows that if one considers higher order effects of $\Delta t$ 
the  optimal acceptance might be slightly changed. In the current study we omit this small effect.} as \cite{Takaishi}
\begin{eqnarray}
\langle P_{acc}\rangle_{opt} & = &   \exp(-\frac{1}{n}) \label{PAopt} \\
& =&  \left\{
\begin{array}{ll}
0.61 & \mbox{2nd} \\
0.78 & \mbox{4th} \\
0.85 & \mbox{6th}.
\end{array}
\label{PAopt2}
\right.
\end{eqnarray}
Note that the above result does not depend on the specific Hamiltonian and 
can be applied for any model. 

Using eq.(\ref{DTopt}) and (\ref{PAopt}) we obtain the optimal efficiency
of the $n$-th order integrator:
\begin{equation}
(E_{ff})_{opt}^{n-th}= \exp(-\frac{1}{n}) \sqrt[n]{\frac{1}{n
\widetilde{C}_n V^{\frac{1}{2}}}  },
\label{Eopt}
\end{equation}
We use eq.(\ref{Eopt}) to compare among integrators.
Eq.(\ref{Eopt}) is easy to handle because  eq.(\ref{Eopt})
has one unknown parameter  $\widetilde{C}_n$  
and the value of $\widetilde{C}_n$ can be estimated easily 
from a single simulation on a rather small lattice. 

Now let us compare $n$-th and $m$-th order integrators $(n>m)$.
If one obtains a gain $G$ for the $n$-th order integrator over the $m$-th order one,
the following condition must be satisfied:
\begin{equation}
(E_{ff})^{n-th}_{opt}= G k_{nm} (E_{ff})^{m-th}_{opt},
\label{comp}
\end{equation}
where $k_{nm}$ is a relative cost factor needed to implement the $n$-th order 
integrator against the $m$-th one and $G$ is defined so that 
both  $n$-th and $m$-th order integrators are equally effective with $G=1$.
In our case, $k_{42}=3$ and $k_{64}=7/3$.
Substituting eq.(\ref{Eopt}) to eq.(\ref{comp})
and rewriting the equation
one obtains the lattice volume needed to have the gain $G$:
\begin{equation}
V^{\frac{n}{2}-\frac{m}{2}} = (G k_{nm}\exp(-\frac{1}{m}+\frac{1}{n}))^{n m}
\left(\frac{1}{m\widetilde{C}_m}\right)^{n} (n\widetilde{C}_n)^m.
\label{comp3}
\end{equation}

Our formulas are based on the assumption that $\langle \Delta H^2\rangle^{1/2}$ 
satisfies eq.(\ref{dH2}).
The validity of  eq.(\ref{dH2}) depends on the action we take and simulation parameters.
In general it is expected that the contribution of the higher order terms to eq.(\ref{dH2})
becomes small as the lattice size increases.
Let us rewrite eq.(\ref{dH2}) as 
\be
\langle \Delta H^2\rangle^{1/2} = C_n V^{1/2} \Delta t^n + C_{n+1} V^{1/2} \Delta t^{n+1},
\ee
where $ C_{n+1} V^{1/2} \Delta t^{n+1}$ stands for the first relevant higher order term.
We measure the contribution of the higher order term by the ratio:
\be
\frac{ C_{n+1} V^{1/2} \Delta t^{n+1}}{ C_n V^{1/2} \Delta t^n}=\frac{ C_{n+1}  }{ C_n } \Delta t.
\label{cratio}
\ee
In order to keep the constant acceptance, $\Delta t$ must be taken to be small  
as the lattice size increases. Thus eq.(\ref{cratio}), ie. the contribution of the higher order term,  
becomes small for larger lattices.  

Fig.1 shows a typical example of $C_n$ on a $16^3\times4$ lattice as a function of $\Delta t$.  
$\langle\Delta H^2\rangle ^{1/2}$ are well expressed by the functions proportional to $\Delta t^n$ except for large $\Delta t$.
We are only interested in the acceptance region indicated by eq.(\ref{PAopt2}): $P_{acc}\geq 60\%$ which corresponds to 
$\langle \Delta H^2\rangle^{1/2}\leq 1$\footnote{See Fig.1 in \cite{Takaishi}.}.
It seems that for $\langle \Delta H^2\rangle^{1/2}\leq 1$
the effect of the higher order terms is small. 

\section{Simulation results}
We use the plaquette gauge action and standard Wilson fermion action
with two flavors of  degenerate quarks ($n_f=2$).
We first determine the coefficients $C_n$ at zero and finite temperature.
This can be done by measuring $\langle \Delta H^2\rangle^{1/2}$ at a small 
step-size and substituting the results into eq.(\ref{dH2}). 
The trajectory length $\tau$ is set to 1.
We choose $\beta=5.75$ and make simulations on both $12^4$ and $12^3\times 4$ lattices.
The critical kappa $\kappa_c$ of $n_f=2$ at this $\beta$
is around 0.157\footnote{This value is estimated from Figure 5 in Ref.\cite{Iwasaki}}. 
We use several $\kappa$'s in a range of $0.1\leq\kappa\leq 0.155$.
In this range we maintain the confinement phase on the $12^4$ lattice and the deconfinement 
phase on the $12^3\times 4$ lattice.
In this study we refer to the results on the  $12^4$($12^3\times 4$) lattice 
as those at zero(finite) temperature. 

Figs.2-4 show $C_n$ as a function of $m_q a$. 
Here $m_q a$ is defined by $\displaystyle m_q a=(1/\kappa-1/\kappa_c)/2$.
At large quark masses where the fermionic effects are negligible,
values of $C_n$ at zero and finite temperature coincide each other. 
This may indicate that the contribution to $C_n$ from the gauge sector 
is almost same at zero and finite temperature. 
At small quark masses $C_n$ at zero temperature increases more rapidly 
than those at finite temperature as $m_q a$ decreases.
The quark mass dependence of $C_n$ at finite temperature, compared to that at zero temperature, 
is small for all the integrators studied here.
This behavior is consistent with the result of Ref.\cite{Accept}.
Substituting values of $C_n$ at finite temperature into eq.(\ref{comp3}) with  $G=1$,
we calculate the lattice size $L_e$ ( here $V=L_e^3\times N_t$ ).
This lattice size $L_e$ is the one with which 
the higher order integrator and the lower order one perform equally.  
For a lattice size $L>L_e$ the higher order integrator is more effective than the lower one.
Fig.5 shows $L_e$ from comparison between the 2nd and 4th order integrators (2nd vs. 4th) 
and between the 4th and 6th order ones (4th vs. 6th).  
For the case of 4th vs. 6th, $L_e$ increases as $m_q a$ decreases, which we do not appreciate.
On the other hand for the 2nd vs. 4th, $L_e$ remains less than 20 even at small $m_q a$.
This result is contrast to that obtained at zero temperature 
where $L_e$ increases as $m_q a$ decreases \cite{Takaishi}.
The above result encourages us to use the 4th order integrator at finite temperature.

The results in Fig.5, however, just show the lattice on 
which the both integrators are equally effective. 
To use the higher order integrator in simulations one must obtain some gain over the lower order one.
In Fig.6, using eq.(\ref{comp3}), we show the expected gain ( the  region between the solid lines ) 
at $\kappa = 0.1525(m_qa\approx 0.094)$ as a  function of lattice size $L$.
To have $G=2$ gain ( which means 2 times faster ) a lattice size $L\approx 100$ is required.
This huge lattice size is still not accessible in the current large-scale simulations.
Probably the maximum lattice size accessible at the moment is $L\leq50$.
Therefore we can not expect a large gain from the 4th order integrator even if it is used now. 
If we use a lattice with $L\approx50$, $G\approx1.5$ can be achieved. 
Thus at the level of the current large-scale simulations, we expect to obtain $G\leq 1.5$. 

We also make simulations at $\kappa = 0.1525$ 
to confirm that we can actually obtain some gain for the 4th order integrator over the 2nd one. 
At $\kappa = 0.1525$, $L_e$ is estimated to be 17.5(9).
We choose $18^3\times 4$ and $28^3\times 4$ lattices.
On the $18^3\times 4$ lattice we expect $G\approx1$ and on the $28^3\times 4$ lattice, $G>1$.  
The step-size is adjusted so that the acceptance gives a similar value with eq.(\ref{PAopt2}).
The gain $G$ is calculated by 
\be
G=\frac{(P_{acc}\times \Delta t)_{4th}}{3(P_{acc}\times \Delta t)_{2nd}},
\ee
where a factor of 3 in the denominator comes from the relative cost factor $k_{42}=3$. 
Table 2 shows the simulation results. 
Using these results, we obtain $G=1.08(3)$ on the $18^3\times 4$ lattice 
and $G=1.35(8)$ on the $28^3\times 4$ lattice (see also Fig.6). 
As expected the gain increases with $L$.
The result on the $28^3\times 4$ lattice 
is an example showing that the 4th order integrator is more effective   
than the 2nd order one.


\begin{table}
\begin{center}
\begin{tabular}{c|cc|cc}
                  & \multicolumn{2}{c|}{$18^3\times 4$} & \multicolumn{2}{c}{$28^3\times 4$}  \\
 $\kappa=0.1525$  &  2nd & 4th   &  2nd & 4th  \\ \hline
$\Delta t$ &  1/24  & 1/10       & 1/36      & 1/12 \\
Acceptance       &  0.57(2)   & 0.81(2)    & 0.60(3)     & 0.81(3) \\
\end{tabular}
\caption{\small Step-size and acceptance at $\kappa=0.1525$.}
\end{center}
\end{table}

\section{Conclusions}
We have studied higher order integrators for the HMC algorithm 
with the Wilson fermion action at finite temperature.
Contrast to the zero temperature case, the 4th order integrator at finite temperature 
can be more effective than the 2nd order one. 
This was demonstrated by the simulations at $\beta=5.75$
on $L^3 \times 4$ lattices. 
The gain is dependent of the lattice size. 
It was shown that on the $28^3\times 4$ lattice
at $\beta=5.75$ and $\kappa=0.1525$ the 4th order integrator
is about $35\%$ faster than the 2nd order one.

When large-scale simulations at finite temperature are planned,
it is recommended to check which integrator is effective for the lattice considered. 
This check can be done easily. 
First we measure $C_2$ and $C_4$.  
This first step does not take much computational time 
since they can be measured on a small lattice.
Then substituting values of $C_2$ and  $C_4$ to eq.(\ref{comp3}), we obtain a relation between $G$ and $L$. 
If we obtain $G>1$ on the lattice planned for the simulations,
we should consider to use the 4th order integrator.

Eq.(\ref{comp3}) is obtained by using the approximation at small $\Delta t$. 
There might exist the systematic errors caused by the approximation.
Although for the present study we considered the errors to be small, for other actions 
and simulation parameters they might contribute largely.
We have to keep in mind that there might exist such contributions depending on the simulation details.

\section*{Acknowledgments}
The simulations were done on the NEC SX-5 at INSAM Hiroshima University and at 
Yukawa Institute. 
The author would like to thank Atsushi Nakamura for 
useful discussion and comments.
This work was supported by the Grant in Aid for Scientific Research by
the Ministry of Education, Culture, Sports, Science and Technology(No.13740164).

\begin{figure}[ht]
\centerline{\psfig{figure=  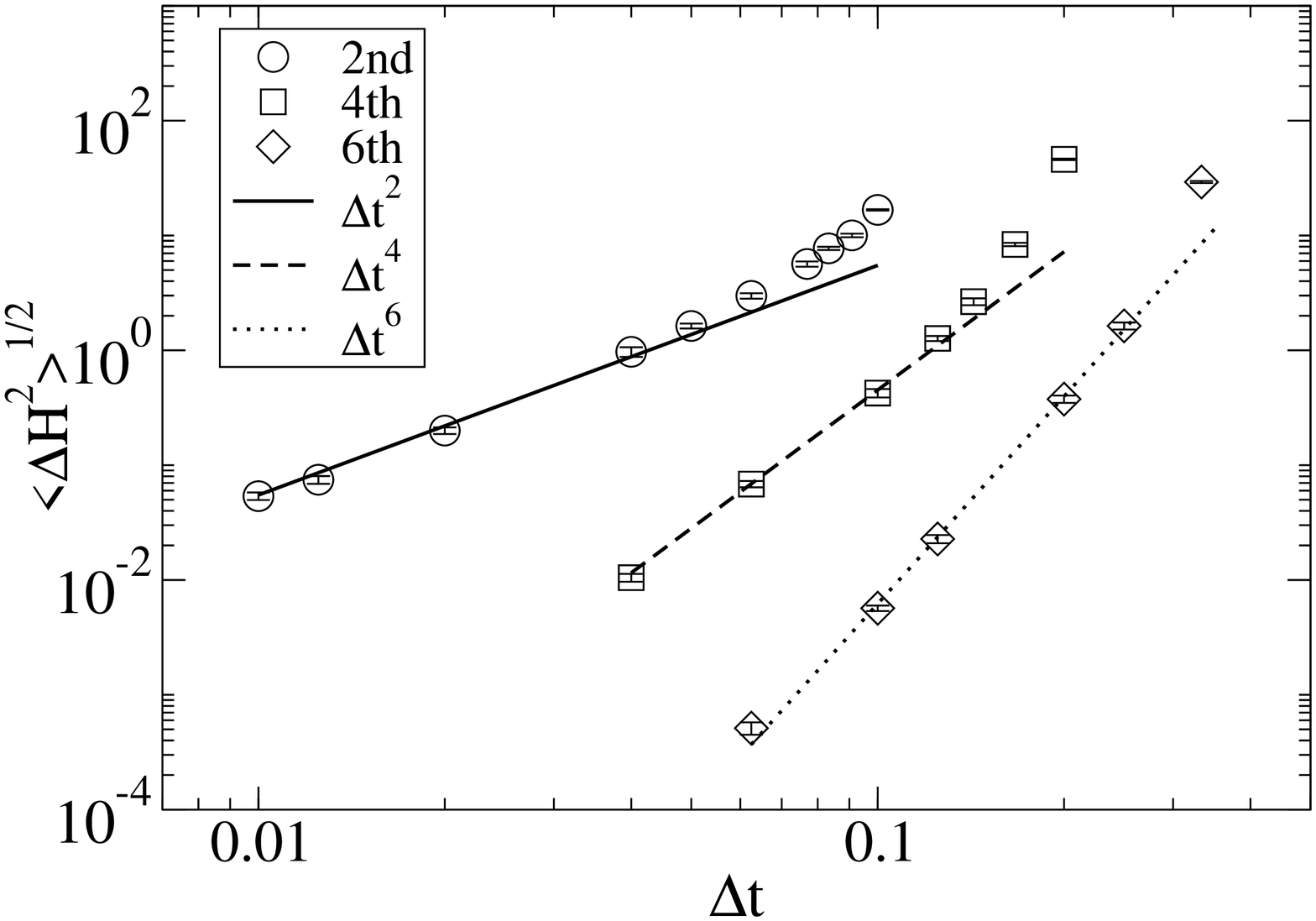,height=8cm}}
\caption{
$\langle \Delta H^2\rangle^{1/2}$  on a $16^3\times 4$ lattice at $\beta=5.75$ and $\kappa=0.155$ 
as a function of $\Delta t$. The lines proportional to $\Delta t^n$ are also drawn.
The simulations were done with the Wilson fermion action.
}
\end{figure}

\begin{figure}[ht]
\centerline{\psfig{figure=  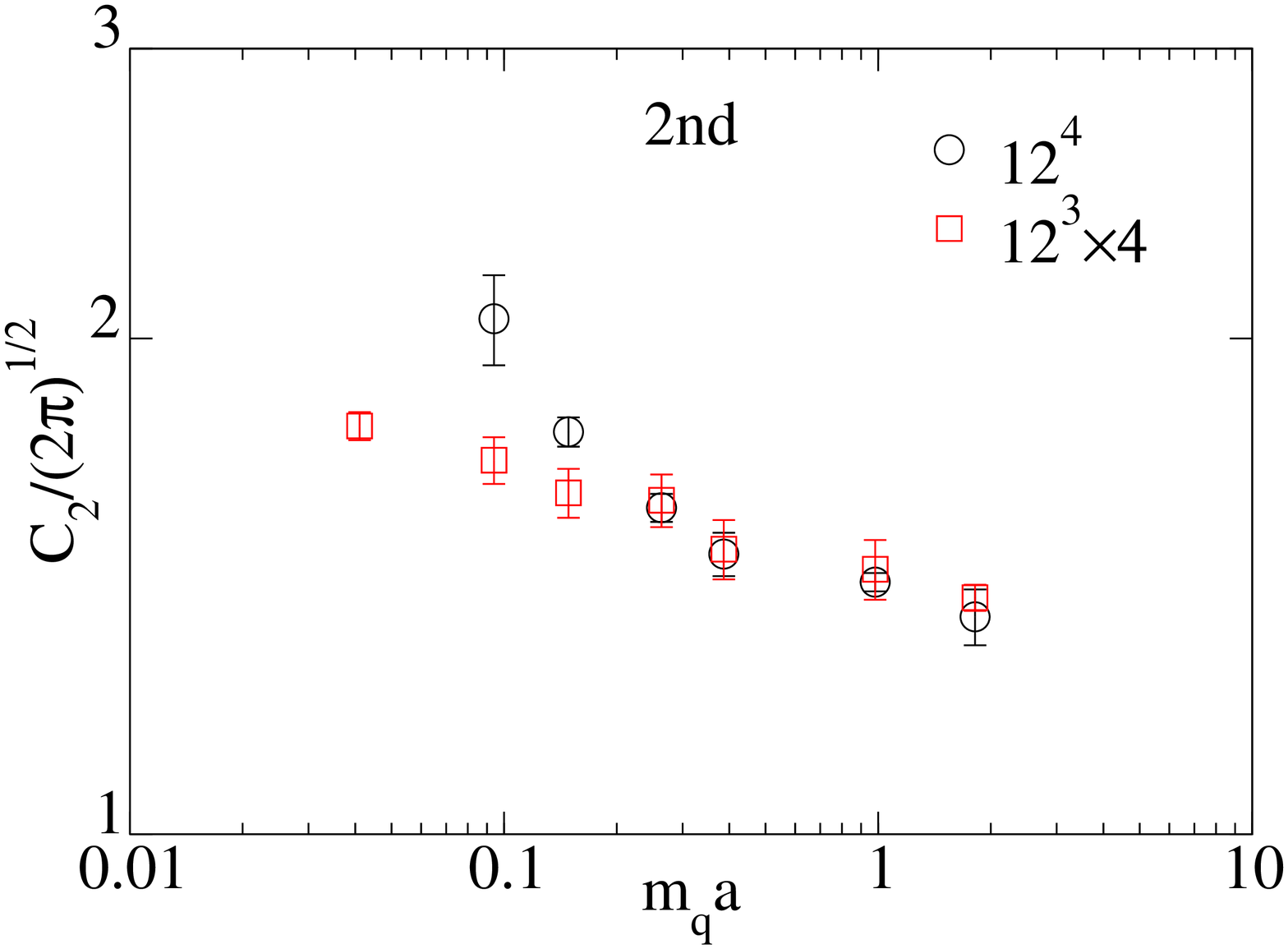,height=8cm}}
\caption{
$C_2$ as a function of the quark mass $m_q a$.
In the figure, $C_2$ is normalized as $C_2/(2\pi)^{1/2}\equiv \tilde{C_2}$.
}
\end{figure}

\begin{figure}[ht]
\centerline{\psfig{figure=  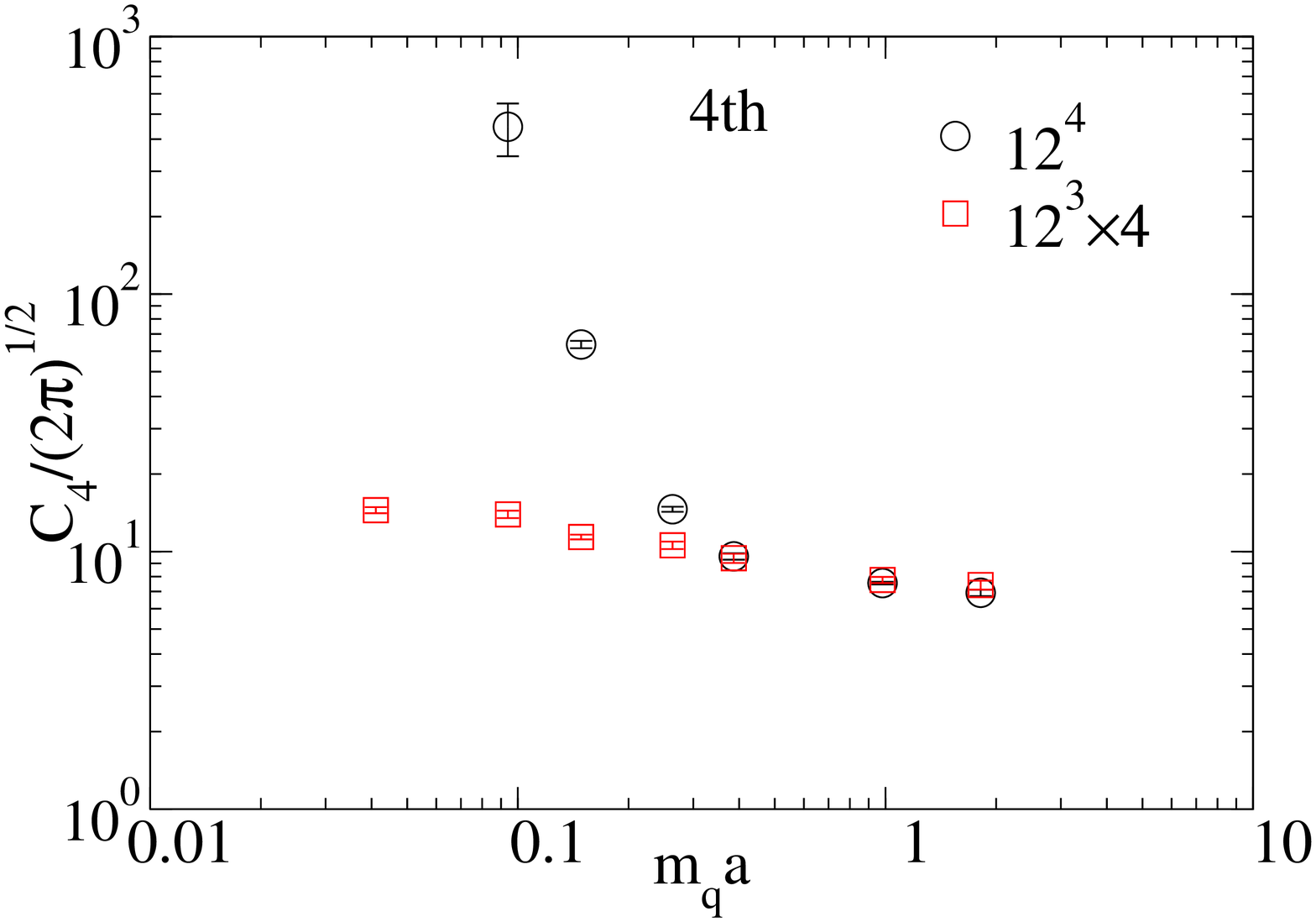,height=8cm}}
\caption{
Same as in Fig.2 but for $C_4$.
}
\end{figure}

\begin{figure}[ht]
\centerline{\psfig{figure=  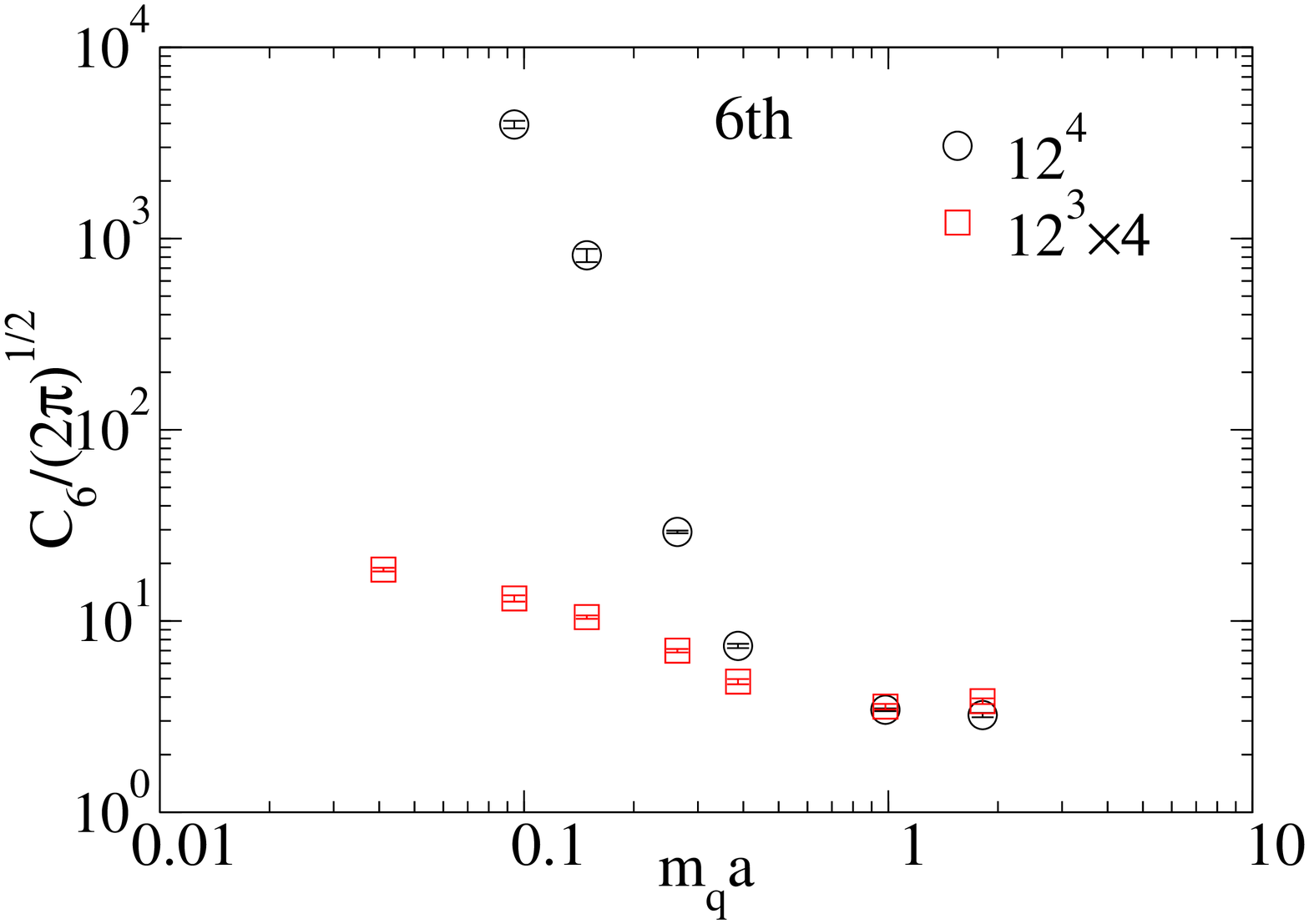,height=8cm}}
\caption{
Same as in Fig.2 but for $C_6$.
}
\end{figure}

\begin{figure}[ht]
\centerline{\psfig{figure=  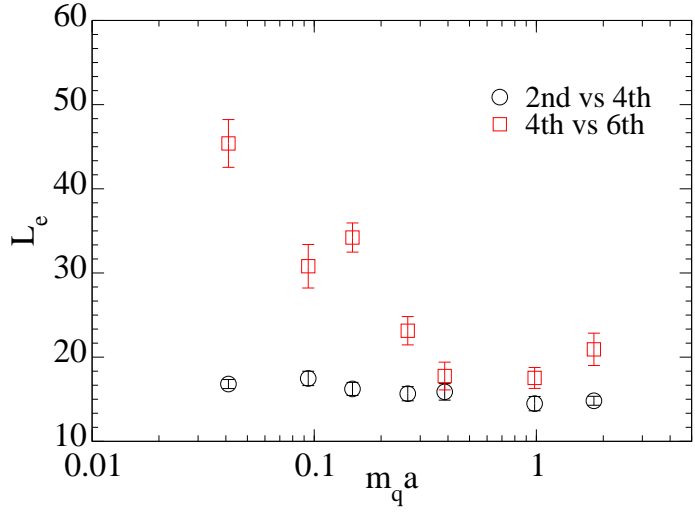,height=8cm}}
\caption{
$L_e$ as a function of the quark mass $m_q a$. 
$L_e$ stands for the lattice size for which the higher order integrator and
the lower order one are equally effective. 
Circles are from comparison between the 2nd order integrator and the 4th order one,
and squares are  comparison between the 4th order integrator and the 6th order one.
}
\end{figure}

\begin{figure}[ht]
\begin{flushright}
\centerline{\psfig{figure=  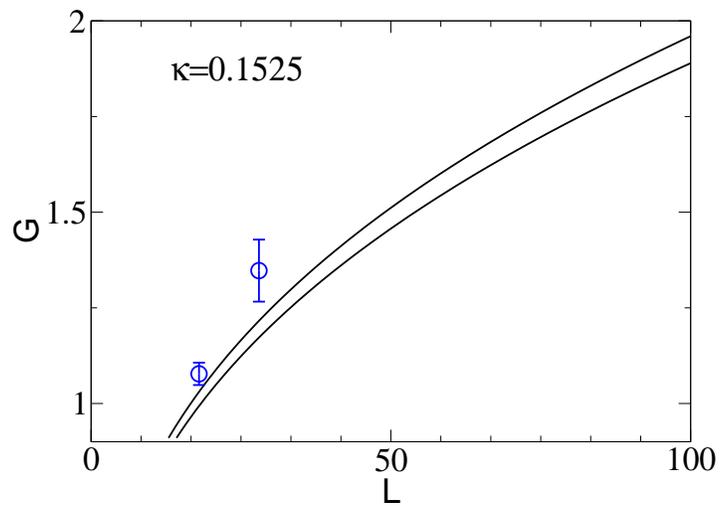,height=8cm}}
\end{flushright}
\caption{
The gain $G$ as a function of $L$ ( here $V = L^3\times 4$ ) at $\kappa=0.1525$.
The solid lines are calculated from eq.(\ref{comp3}).
Circles are from Monte Carlo simulations.
}
\end{figure}

\end{document}